\documentclass[preprint,12pt, a4paper]{article}

\usepackage{amssymb}

\usepackage{lineno}

\usepackage{float}
\restylefloat{table}
\usepackage{graphicx}

\usepackage{hyperref}
\usepackage{listings}

\renewcommand{\lstinline}[1]{{\bfseries\oldlstinline{#1}}}
\newcommand{\keywords}[1]{\textbf{Keywords:} #1}
\usepackage[table]{xcolor}

\begin{document}

\title{\textit{scg-cli} -- a Tool Supporting Software Comprehension via Extraction and Analysis of Semantic Code Graph} 

\author{
  Krzysztof Borowski\thanks{AGH University of Krakow and VirtusLab, Poland (e-mail: kborowski@agh.edu.pl, kborowski@virtuslab.com)} \and 
  Bartosz Balis\thanks{Institute of Computer Science, AGH University of Krakow, Krakow, Poland (e-mail: balis@agh.edu.pl)}  
}

\date{}
\maketitle

\begin{abstract}
We present \textit{scg-cli}, a~command line tool facilitating software comprehension. The tool extracts semantic information about code structure and dependencies from the Java and Scala projects, and structures it as a~Semantic Code Graph (SCG), an information model underlying \textit{scg-cli}. The SCG data, once written into a~portable, open protobuf-based format, can be used by the \textit{scg-cli} command line tool to obtain project metrics, find the most critical code entities, and compute project partitionings. The results of this analysis and the SCG data can be exported for further investigation by external tools such as Gephi software (visualization) and, notably, as a Jupyter Notebook environment with helper APIs to enable advanced analysis of the project using data analytics methods. We explain functionalities of the \textit{scg-cli} tool and demonstrate its capabilities by showing an example analysis of an open-source Java project \textit{commons-io}. 
\end{abstract}


\section{Motivation and significance}

Every day, programmers write new lines of code, fueling the growth of software repositories. Once written, the code undergoes multiple rounds of reading and modification, often not by the original author. As a result, software maintainers are faced with the task of comprehending unfamiliar sections of the software \cite{Al-Saiyd2017}. In addition, developers widely reuse open-source libraries, which often entails the need for a deep understanding of the library's source code, not only to enable proper usage but also to introduce bug fixes and new features.

Understanding unfamiliar software, whether it is an in-house project or a third-party open-source library, can be an arduous and time-consuming undertaking for developers \cite{Xia2018}. They strive to gain comprehensive knowledge of the project as quickly as possible, however, while project documentation typically offers instructions on building and preparing merge requests, there is often a lack of architecture documentation that provides a holistic view and introduction to the project. Consequently, critical information about the software needs to be extracted directly from the source code. The \textit{scg-cli} aims to accelerate the software comprehension \cite{Siegmund2016, Schroter2017} process by providing insights extracted directly from the source code, expediting the learning curve for developers.

The \textit{scg-cli}\footnote{\url{https://github.com/virtuslab/scg-cli}} utilizes the Semantic Code Graph (SCG) \cite{Borowski2022}, an information model closely tied to the source code, to capture software artifacts and code dependencies. By leveraging this model, the \textit{scg-cli} can generate SCG data from Java projects and employ various software comprehension functionalities based on this data. Additionally, it offers the potential for extending analysis to other languages either through enhancing the capabilities of \textit{scg-cli} or by utilizing external SCG data extractors, such as the separate compiler plugin available for Scala\footnote{\url{https://github.com/VirtusLab/scg-scala}}. To ensure compatibility, integrations for other languages must adhere to the official metadata structure defined in the protobuf format\footnote{\url{https://github.com/VirtusLab/graphbuddy/blob/master/proto/graph\_node.proto}}.

The \textit{scg-cli} command line tool enables software comprehension of projects in several aspects. Currently, its capabilities include:
\begin{itemize}
    \item Generating project summary to give the programmer high level overview of the project.
    \item Finding critical entities in the source code that are usually good starting points for reading the project.
    \item Discovering and suggesting software partitioning to expose module boundaries or support refactoring efforts.
    \item Exporting SCG data and analysis results to graph formats (.GDF, graphml) for external visualizations.
    \item Providing a Jupyter-based environment for further software mining through analysis of SCG data using popular scientific libraries, such as pandas\footnote{\url{https://pandas.pydata.org/}}, or networkx\footnote{\url{https://networkx.org/}}.
\end{itemize}

The functionalities of the \textit{scg-cli} significantly reduce the time needed to comprehend software by extracting and presenting invaluable project-specific knowledge. Furthermore, it establishes a robust foundation for future functionalities, tools, and extensions developed based on the SCG model of the source code.

In software development, the activities of software comprehension and maintenance are typically performed within integrated development environments (IDEs). IDEs primarily enable efficient browsing and manipulation of source code files. Browsing the source code is a key element in software comprehension, and within IDEs, it is augmented with syntax highlighting, go-to capabilities, file hierarchy, and multiple other facilitators. However, IDE-based software comprehension is focused on the particular context with a limited scope of the source code unit. For advanced and comprehensive software analysis there exist dedicated tools such as \textit{JArchitect}\footnote{\url{https://www.jarchitect.com/}}, or specialized IDE plugins, such as \textit{stan4j}\footnote{\url{http://stan4j.com/intro/}}. \textit{JArchitect} is a standalone tool capable of monitoring code quality, comparing releases, or finding and prioritizing technical debt for refactoring activities. It also provides project overview through exploring software architecture capabilities or trends monitoring. However, its software comprehension capabilities are quite limited, e.g., it lacks the ability to find crucial code entities. Moreover, the tool is not open source and hence impossible to extend, it is restricted to a paid license and only supports the Java language. \textit{stan4j} analyzes code structure quality based on the generated Java bytecode. It has an Eclipse integration plugin to enable code structure visualization and browsing, which can help with software comprehension capabilities. On the other hand, it is a paid tool for bigger projects (the free version is limited to 500 classes), and does not allow for custom analysis. The tool is also restricted to Java support and it has not been actively maintained since 2021. There are multiple widely used tools available for static code analysis focusing on code review, security or code quality, such as ScalaStyle\footnote{\url{http://www.scalastyle.org/}}, Checkstyle\footnote{\url{https://checkstyle.org/}}, or monitoring platforms such as SonarQube\footnote{\url{https://sonarqube.org}} or Fortify\footnote{\url{https://www.microfocus.com/en-us/cyberres/application-security}}. These tools, however, are not suitable for supporting the software comprehension process. In contrast, \textit{scg-cli} focuses on software comprehension in the context of the entire project, possibly narrowing down the investigation scope if needed.

\section{Software description}

\subsection{Software Architecture}

The \textit{scg-cli} is a command line application written in Scala 3 and it can be built directly from the sources using the \textit{sbt}\footnote{\url{https://www.scala-sbt.org/}} \lstinline{sbt universal:packageBin} command, but for user convenience binary artifacts are distributed through the official Github page\footnote{\url{https://github.com/VirtusLab/scg-cli/releases}}. A~compressed package contains compiled source code with run scripts for MacOS, Linux and Windows operating systems. For basic usage, the tool requires only a~Java runtime preinstalled on the host machine. 

\begin{figure}[!htbp]
\includegraphics[width=12cm]{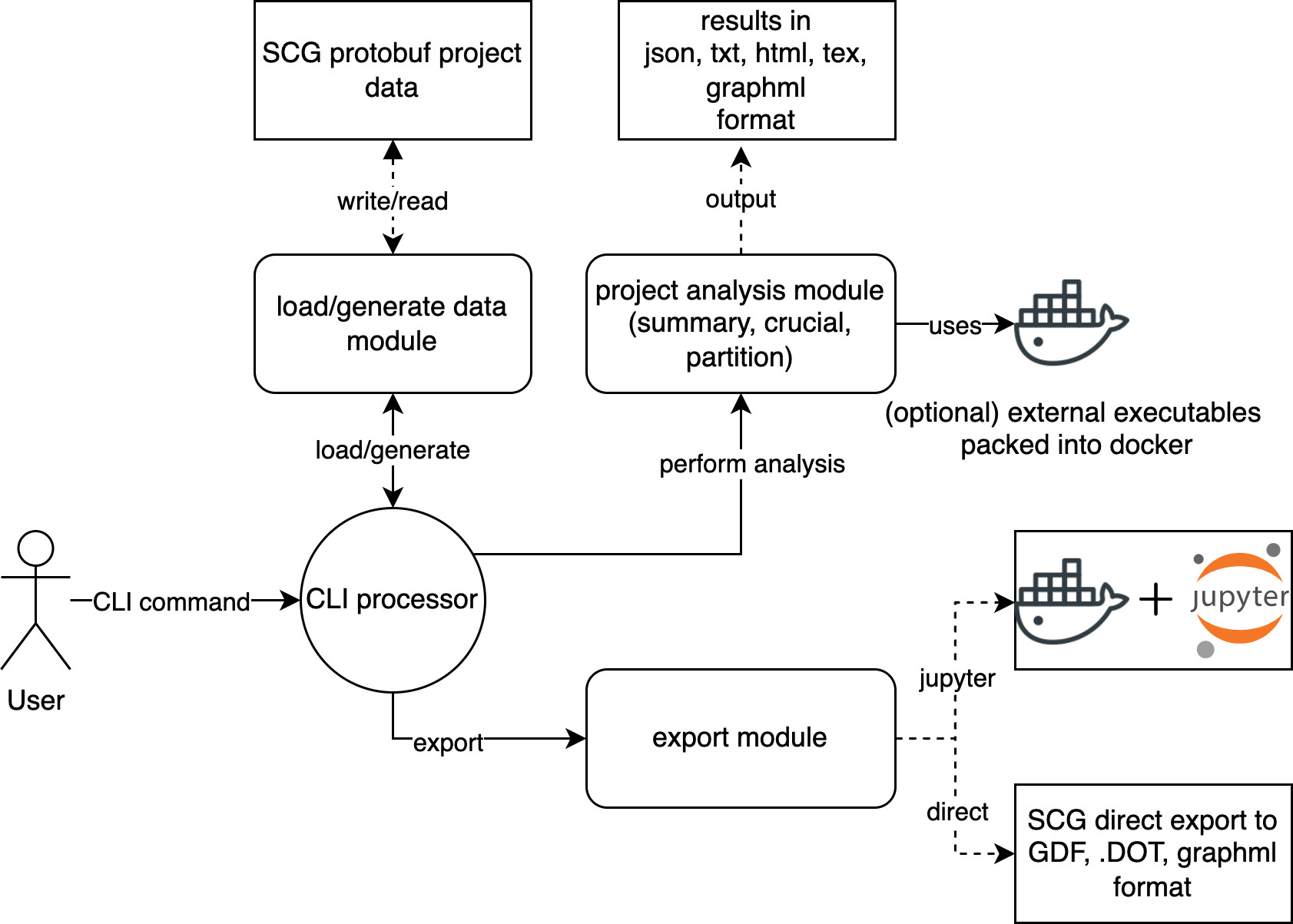}
\caption{The \textit{scg-cli} software architecture. The \textit{load/generate data module} generates SCG data files for a~project. The \textit{project analysis module} performs analysis of the project using the SCG data. The \textit{export module} enables custom analysis using external tools known to researchers, such as Jupyter.}
\label{fig:scg-cli-architecture}
\end{figure}

The architecture diagram for the \textit{scg-cli} tool is presented in Fig. \ref{fig:scg-cli-architecture}. All interactions with the tool are invoked by user commands and processed by the CLI processor. Three different modules can be involved in the process of executing user commands:
\begin{itemize}
    \item load/generate data module -- responsible for initial source code analysis and generating SCG protobuf metadata files or loading the SCG data, 
    \item project analysis module -- conducting various project analyzes based on the SCG information model,
    \item export module -- exporting SCG data to other formats (e.g. GDF, .graphml) for further analysis with external tools such Gephi\footnote{\url{https://gephi.org/}} or exporting and running a Jupyter notebook \footnote{\url{https://jupyter.org/}} with loaded SCG data.
\end{itemize}

The capabilities of \textit{scg-cli} can be leveraged by external software analysis tools by utilizing directly its command line interface. This integration pattern is widely embraced in the industry for interacting with third-party software.

\subsection{Software Functionalities}
\label{sec:software-functionalities}

The \textit{scg-cli} implements helper command \lstinline{scg-cli help} which summarizes software functionalities, as presented below.
\begin{small}
\begin{verbatim}
Usage: scg-cli [COMMAND]
CLI to analyze projects based on SCG data
Commands:
  help       Display help information about the specified 
             command.
  crucial    Find crucial code entities.
  generate   Generate SCG metadata.
  partition  Suggest project partitioning.
  summary    Summarize the project.
  export     Export SCG metadata to various output formats. 
\end{verbatim}
\end{small}
Brief information about the usage of each particular command can be obtained with \lstinline{scg-cli help [COMMAND]}. In the next subsections we will explain in detail major \textit{scg-cli} capabilities.

\subsubsection{Generating SCG data}

The \textit{scg-cli} tool is able to generate \textit{SCG} protobuf metadata for Java projects using command \lstinline{scg-cli generate -l java <workspace>}. Generated data is stored in the project structure under \texttt{.semanticgraphs} folder. It is also possible to generate \textit{SCG} data for Scala projects with the help of the Scala plugin\footnote{\url{https://central.sonatype.com/namespace/org.virtuslab.semanticgraphs}}. These files are later loaded into the \textit{scg-cli} for further analysis. 

\subsubsection{Presenting project overview}

Project overview allows one to grasp the size of the project and other interesting characteristics. The \lstinline{scg-cli summary <workspace>} command produces short project overview containing SCG metrics, such as the number of nodes and edges and their distribution by type, total number of code lines, graph density, average node in- and out-degree, global clustering coefficient, degree assortativity coefficient. Each metric provides some insight about the project, e.g.:
\begin{itemize}
    \item Number of nodes and edges -- a simple metric to asses the project size and monitor its growth over the time.
    \item Density -- high SCG density can indicate higher code dependency coupling and general complexity.
    \item Global Clustering Coefficient -- high value can indicate well-modularized software with high module cohesion and low coupling.
    \item Node type distribution -- provides various insights, e.g., a high number of variables can entail harder reasoning about the code and overall lower software quality.
\end{itemize}

\subsubsection{Finding crucial code entities}
\label{subsec:f-crucial}

The most important central code entities are typically good starting points for reading the source code in the process of software comprehension \cite{Du2021}. The \textit{scg-cli} automatically finds the most important nodes in the SCG graph based on the set of nine different metrics \cite{Gomez2019}, including LOC (line of codes associated with a~code entity), out-degree and in-degree, eigenvector centrality, Katz centrality, PageRank, betweenness centrality and Harmonic Centrality. Each metric represents the importance of a~node from a~different perspective. We also introduced an additional metric -- \textit{Combined importance} -- which represents node importance aggregating all other metrics, i.e., a~node with high score in multiple metrics will also have a~high combined importance.

\subsubsection{Suggesting project partitioning}

Software Module Clustering is a data mining technique that groups software entities into clusters sharing same features \cite{Sarhan2022}, facilitating software comprehension \cite{Mohammadi2019} or refactoring activities, e.g., software modularization. The \textit{scg-cli} tool provides clustering functionality which uses graph partitioning algorithms \cite{bulucc2016recent} to divide the SCG graph into a~certain number of partitions while minimizing, e.g., the number of edges that cross partitions. The maximum number of the partitions is provided as an command line argument \texttt{n}:\\ \lstinline{scg-cli partition -o \{csv,json,html,txt,tex,gml\} <workspace> \{n\}}.

As the SCG graph often contains thousands of nodes, the \textit{scg-cli} tool uses two efficient graph partitioning algorithms implemented in gpmetis \cite{karypis1997metis} and patoh \cite{Ccatalyurek2011} software. Both algorithms compute partitions within seconds even for large graphs.

The partitioning functionality will not only present the node allocation to each partition but also will compare the quality and characteristics between different partitionings. The following properties of the partitioned graphs are compared:
\begin{itemize}
    \item Modularity -- the ratio of edges between nodes inside a~partition to edges crossing the partition boundary.
    \item Average Clustering Coefficient -- the measure of tendency of nodes to cluster together in the partition, computed separately for each partition and averaged.
    \item Weighted and Average Accuracy by file and by package unit -- average percentage of nodes within a~given file or package unit which belongs to the same partition. For a~well structured software all nodes in a~particular file or package should be assigned by the algorithm to the same partition. Outstanding nodes might imply the need to reorganize the source code and improve the project structure quality. Weighted Accuracy additionally takes into account the size (in term of the total number of nodes) of a~particular file or package.
    \item Partition variance -- a~measure of partition size similarity. A value close to zero means that the number of nodes in all partitions are equal, while a~value close to 1 denotes highly unbalanced partitions in terms of size.
    \item Nodes distribution -- a~summary which presents the number of nodes within each partition (as percentage of all nodes).
\end{itemize}

\subsubsection{Exporting for further analysis in other tools}

The \textit{scg-cli} can export the data to external format, i.e. GDF\footnote{\url{https://gephi.org/users/supported-graph-formats/gdf-format/}}, .DOT\footnote{\url{https://en.wikipedia.org/wiki/DOT_(graph_description_language)}} or graphml\footnote{\url{https://en.wikipedia.org/wiki/GraphML}} which can be later consumed in other tools such as Gephi\footnote{\url{https://gephi.org/}} or Graphviz\footnote{\url{https://graphviz.org/}}. Moreover, a~special \textit{jupyter} export allows one to run a~Jupyter notebook in a~Docker container and load all SCG data into it for further software mining activities, using command \lstinline{scg-cli export -o jupyter <workspace>}. The command mounts the SCG data as a Docker volume and starts a~publicly available Docker image\footnote{\url{https://hub.docker.com/r/liosedhel/scg-jupyter}} containing a~Jupyter notebook along with helper API to load and operate over SCG data. After successful container startup the resulting local Jupyter notebook URL is presented to the console.

\section{Illustrative Examples}
\label{illustrative-example}

In this section, we demonstrate several software analyzes conducted with the \textit{scg-cli} tool, using the commons-io\footnote{\url{https://github.com/apache/commons-io}} project as an example. The \textit{commons-io} project is an open-source library widely used in the industry to facilitate various file and stream related functionalities\footnote{\url{https://commons.apache.org/proper/commons-io/}} and it is a part of commonly known Apache Commons set of libraries\footnote{\url{https://commons.apache.org/}}.

\subsection{Generating SCG data}

To begin the analysis, the project sources have to be downloaded to the local disk. The \textit{scg-cli} will analyze local files and generate the SCG data using command \lstinline{scg-cli generate -l java commons-io}. The program will parse source files and run detailed analysis to extract a~complete SCG model for all Java sources and store resulting protobuf files into the \texttt{.semanticgraphs} folder. All further analysis using \textit{scg-cli} are performed based on the generated metadata without the need to analyze the source code files again.

\subsection{Project overview}

\begin{figure}[!htbp]
\includegraphics[width=13.5cm]{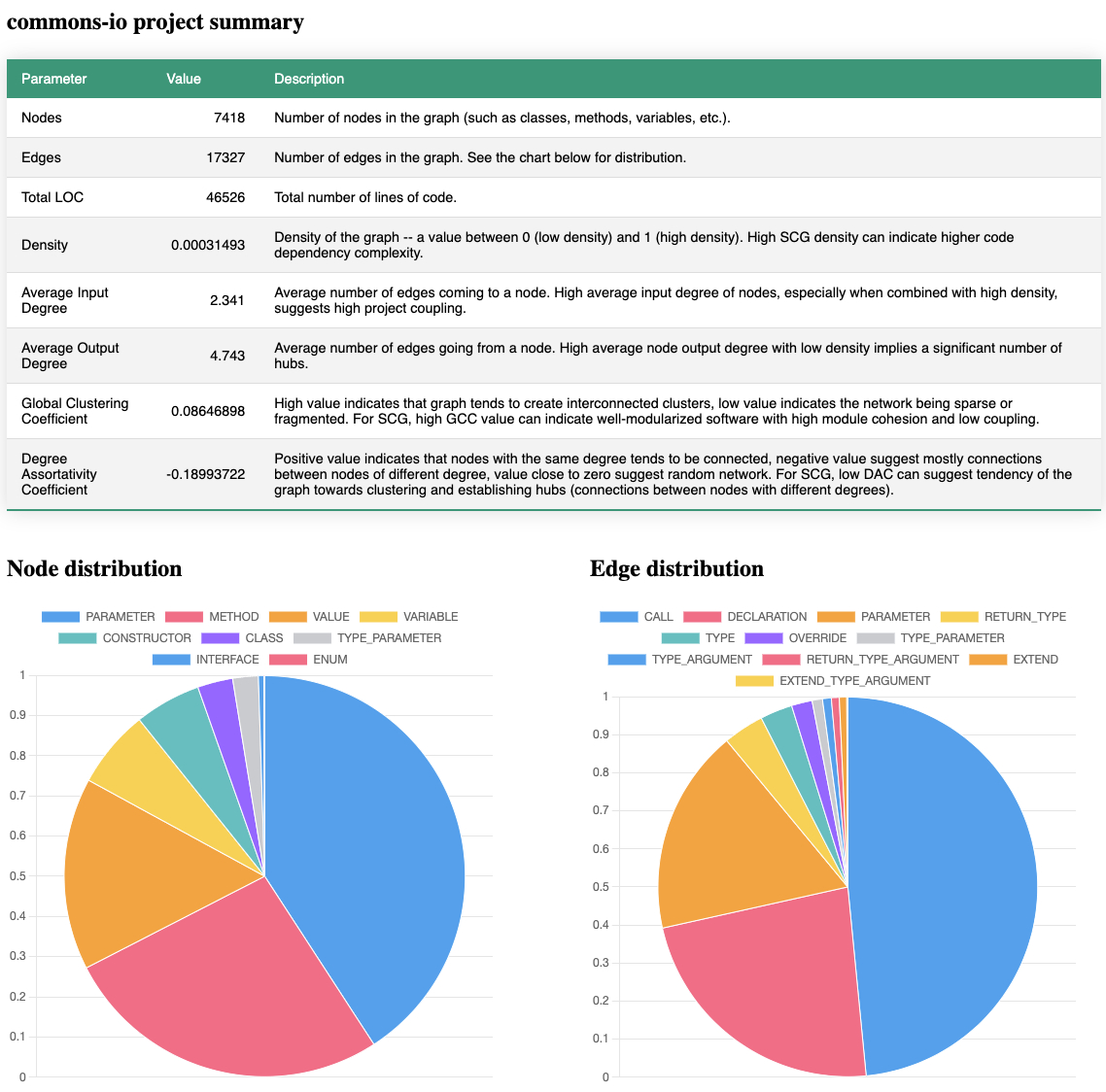}
\caption{Project overview for \textit{commons-io} project generated with \lstinline{scg-cli summary} command.}
\label{fig:commons-io-summary}
\end{figure}

To generate the project overview in the HTML format, one needs to run \lstinline{scg-cli summary -o html commons-io}.  The result, presented in Fig. \ref{fig:commons-io-summary}, contains computed graph parameters and a~chart with distribution of node and edge types. The parameters are explained in the presented table.

\subsection{Finding crucial nodes}

\textit{Finding crucial nodes} is a \textit{scg-cli} feature that analyzes the entire project and applies different scoring metrics to asses importance of graph nodes (i.e., code entities) from different perspectives (the different metrics are described in Section \ref{subsec:f-crucial}). Table \ref{tab:crucial_nodes} presents top 3 nodes in each category computed for the \textit{commons-io} project, obtained from the output of command \lstinline{scg-cli crucial -n 3 -o tex commons-io}. Looking at the combined importance metric, the most important nodes are \lstinline{IOUtils} (utility class which enhances operations on streams) and \lstinline{FileUtils} (utility class which enhances operations on files). These classes indeed contain the most desirable and useful \textit{commons-io} library functionalities. \textit{IOFileFilter} is a trait with 31 different implementations facilitating file filtering, complementing \lstinline{FileUtils} functionalities.

\begin{table}[!htbp]
    \centering
    \begin{footnotesize}
    \begin{tabular}{l|l}
        \hline
        \rowcolor{gray!30}
        Lines of code & Score \\
        \hline
        o.a.c.i.IOUtils & 3608 \\
        o.a.c.i.FileUtils & 3434 \\
        o.a.c.i.file.PathUtils & 1682 \\
        \hline
        \rowcolor{gray!30}
        Node out-degree & Score \\
        \hline
        o.a.c.i.FileUtils & 176 \\
        o.a.c.i.IOUtils & 171 \\
        o.a.c.i.file.PathUtils & 106 \\
        \hline
        \rowcolor{gray!30}
        Node in-degree & Score \\
        \hline
        o.a.c.i.filefilter.IOFileFilter & 131 \\
        o.a.c.i.CloseableURLConnection?urlConnection & 47 \\
        o.a.c.i.function.IOBaseStream.unwrap() & 46 \\
        \hline
        \rowcolor{gray!30}
        PageRank & Score \\
        \hline
        o.a.c.i.filefilter.IOFileFilter & 0.013 \\
        o.a.c.i.output.NullPrintStream & 0.011 \\
        o.a.c.i.output.NullPrintStream.NullPrintStream() & 0.006 \\
        \hline
        \rowcolor{gray!30}
        Eigenvector centrality & Score \\
        \hline
        o.a.c.i.function.IOBaseStream.unwrap() & 0.307 \\
        o.a.c.i.input.Tailer & 0.272 \\
        o.a.c.i.function.IOStream.T & 0.201 \\
        \hline
        \rowcolor{gray!30}
        Katz centrality & Score \\
        \hline
        o.a.c.i.filefilter.IOFileFilter & 2.340 \\
        o.a.c.i.CloseableURLConnection?urlConnection & 1.476 \\
        o.a.c.i.function.IOBaseStream.unwrap() & 1.467 \\
        \hline
        \rowcolor{gray!30}
        Betweenness centrality & Score \\
        \hline
        o.a.c.i.function.IOStreams.forAll() & 301648.755 \\
        o.a.c.i.function.IOConsumer & 299069.829 \\
        o.a.c.i.function.IOStream.adapt() & 272351.667 \\
        \hline
        \rowcolor{gray!30}
        Harmonic centrality & Score \\
        \hline
        o.a.c.i.FileUtils & 0.092 \\
        o.a.c.i.IOUtils & 0.070 \\
        o.a.c.i.FileUtils.FileUtils() & 0.064 \\
        \hline
        \rowcolor{gray!30}
        Combined importance & Score \\
        \hline
        o.a.c.i.FileUtils & 3 \\
        o.a.c.i.filefilter.IOFileFilter & 3 \\
        o.a.c.i.IOUtils & 3 \\
    \end{tabular}
    \end{footnotesize}
    \caption{Importance analysis: top 3 most important nodes (code entities) in the \textit{commons-io} library computed based on 9 different metrics.}
    \label{tab:crucial_nodes}
\end{table}

\subsection{Project partitioning}

The \textit{common-io} project consists of just one module. If we were to divide this software into smaller, dedicated libraries, we would need to consider the number of libraries and their boundaries. Partitioning the project can also unveil unique software functionalities or identify whole subsystems, which supports the process of software comprehension. The \textit{scg-cli} tool can suggest partitionings of the project graph and assess their quality. For example, command \lstinline{scg-cli -o tex partition commons-io 10} will compute partitionings of the \textit{commons-io} project starting from 2 partitions up to 10 partitions. The user will be presented with the partitioning summary containing information shown in Table \ref{tab:commons-io-partition-summary}. From these results we can deduce that the proposed project split into 6 partitions (using gpmetis algorithm) presents exceptionally good quality as measured by modularity metric equal to 103.220, average clustering coefficient above 0.180 and very high accuracy (for file weighted and standard accuracy -- 98\% and 99\% respectively). The sizes of the partitions are not evenly distributed with variance above 0.4 and small first and fourth partitions. To look at details of this split, we can export its data to HTML summary: \lstinline{scg-cli -o html partition commons-io 10}. Partition summary presents for each partition split the dominant partition number for each package and file. Exporting to CSV or GML is also possible which enables even further analysis, as shown in the next section.

\begin{table}[!htbp]
    \centering
    \begin{tiny}
\setlength{\tabcolsep}{0.4em}
\begin{tabular}{|r|r|r|r|r|r|r|r|r|l|} 
\hline
Algorithm &  NPart & Modularity & ACC & F. W. & F. A. & P. W. & P. A. & Variance & Distribution \% \\ 
\hline 
gpmetis & 2 & 1490.154 & 0.117 & 99 & 99 & 95 & 98 & 0.844 & [4,95] \\ 
\rowcolor{gray!30}
patoh & 2 & 100.492 & 0.107 & 95 & 99 & 79 & 91 & 0.024 & [42,57] \\ 
gpmetis & 3 & 39.385 & 0.162 & 88 & 97 & 71 & 84 & 0.450 & [15,19,64] \\ 
\rowcolor{gray!30}
patoh & 3 & 79.436 & 0.148 & 95 & 99 & 75 & 90 & 0.026 & [25,37,37] \\ 
gpmetis & 4 & 67.741 & 0.190 & 95 & 98 & 65 & 83 & 0.372 & [11,26,12,49] \\ 
\rowcolor{gray!30}
patoh & 4 & 58.463 & 0.176 & 91 & 98 & 63 & 83 & 0.047 & [20,21,23,34] \\ 
gpmetis & 5 & 62.143 & 0.201 & 95 & 97 & 59 & 82 & 0.547 & [4,9,10,38,37] \\ 
\rowcolor{gray!30}
patoh & 5 & 41.792 & 0.181 & 90 & 98 & 56 & 80 & 0.059 & [16,16,15,21,28] \\ 
\textbf{gpmetis} & \textbf{6} & \textbf{103.220} & \textbf{0.181} & \textbf{98} & \textbf{99} & \textbf{61} & \textbf{79} & \textbf{0.412} & \textbf{[1,28,22,2,22,23]} \\ 
\rowcolor{gray!30}
patoh & 6 & 35.166 & 0.184 & 88 & 97 & 53 & 79 & 0.044 & [12,14,18,13,18,22] \\ 
gpmetis & 7 & 88.332 & 0.204 & 97 & 99 & 62 & 83 & 0.186 & [5,10,26,12,10,15,18] \\ 
\rowcolor{gray!30}
patoh & 7 & 32.596 & 0.192 & 88 & 98 & 51 & 77 & 0.038 & [10,15,11,14,12,15,19] \\ 
gpmetis & 8 & 69.491 & 0.206 & 97 & 98 & 61 & 83 & 0.038 & [12,10,10,14,11,12,9,17] \\ 
\rowcolor{gray!30}
patoh & 8 & 30.418 & 0.196 & 85 & 95 & 53 & 79 & 0.027 & [9,11,10,13,13,15,11,14] \\ 
gpmetis & 9 & 60.540 & 0.213 & 96 & 98 & 58 & 80 & 0.224 & [7,20,11,5,8,7,6,19,13] \\ 
\rowcolor{gray!30}
patoh & 9 & 23.262 & 0.192 & 82 & 95 & 48 & 72 & 0.022 & [10,10,10,10,11,9,10,15,12] \\ 
gpmetis & 10 & 66.309 & 0.188 & 96 & 98 & 56 & 79 & 0.142 & [11,4,3,7,16,11,9,12,11,11] \\ 
\rowcolor{gray!30}
patoh & 10 & 22.785 & 0.180 & 84 & 96 & 46 & 74 & 0.021 & [9,9,7,9,11,9,11,8,12,10] \\ 
\hline 
\end{tabular}
\end{tiny}
    \caption{Different partitionings of the \textit{commons-io} library and their quality computed by \textit{scg-cli}.}
    \label{tab:commons-io-partition-summary}
\end{table}

\subsection{Exporting and analyzing in external tools}

The \textit{scg-cli} can export the data for further analysis in Jupyter with \lstinline{scg-cli export -o jupyter commons-io}. Here, we would like to analyze further the best partitioning of the \textit{commons-io} library. More specifically, we would like to find methods that belong to a~different partition than their parent class. Classes with such methods can be problematic when splitting the project since they are probably used extensively by different parts of the project.
Listing \ref{lst:outstanding-methods} shows the code that performs this analysis. First, we load the CSV output of partitioning results for partition number 6 (gpmetis) into the notebook (line 4) and merge the results into one data frame (line 5). For each method (line 21) we add two additional columns: \textit{parent} which is a parent identifier and \textit{parent\_npart} which is a parent partition number (line 22). Then we find the methods whose partition number is not the same as the one of their parent classes (line 23) and group by parent identifier to find the number of outstanding methods. Looking at results, we find that when splitting the software to 6 partitions (with gpmetis algorithm), classes \textit{FileUtils} and \textit{IOUtils} have, respectively, 14 and 7 methods that belong to different partitions. Consequently, these classes contain functionalities utilized by different partitions and splitting the software into separate modules will likely require some prior code movements and refactorings.

\begin{lstlisting}[language=Python, basicstyle=\footnotesize\ttfamily, caption={Finding methods assigned to different partitions than their parent class}, label={lst:outstanding-methods},  numbers=left]
import pandas as pd
import scg

scg_files = scg.read_scg("commons-io")
G = scg.create_graph(scg_files)
scg_df = pd.merge(
    scg.create_nodes_df(scg_files),
    pd.read_csv("commons-io-npart-6-gpmetis.csv"),
    on="id",
)
def find_parent_and_parent_npart(m):
    parent = next(
        p
        for p, _, data in G.in_edges(m["id"], data=True)
        if data["type"] == "DECLARATION"
    )
    p = scg_df.query("id == @parent")
    return pd.concat(
        [p["id"], p["npart"]], ignore_index=True
    )
m_df = scg_df.query('kind == "METHOD"').copy()
m_df[["parent", "parent_npart"]] = m_df.apply(
    find_parent_and_parent_npart, axis=1
)
m_df[m_df["npart"] != m_df["parent_npart"]].groupby(
    "parent"
).size()
#o.a.c.io.ByteOrderMark                             1
#o.a.c.io.FileUtils                                14
#o.a.c.io.IOUtils                                   7
#o.a.c.io.comparator.LastModifiedFileComparator     1
#o.a.c.io.file.PathFilter                           1
#o.a.c.io.file.PathUtils                            2
#o.a.c.io.filefilter.EmptyFileFilter                1
#o.a.c.io.input.BOMInputStream                      1
\end{lstlisting}

\begin{figure}[!htbp]
    \centering
    \includegraphics[width=12cm]{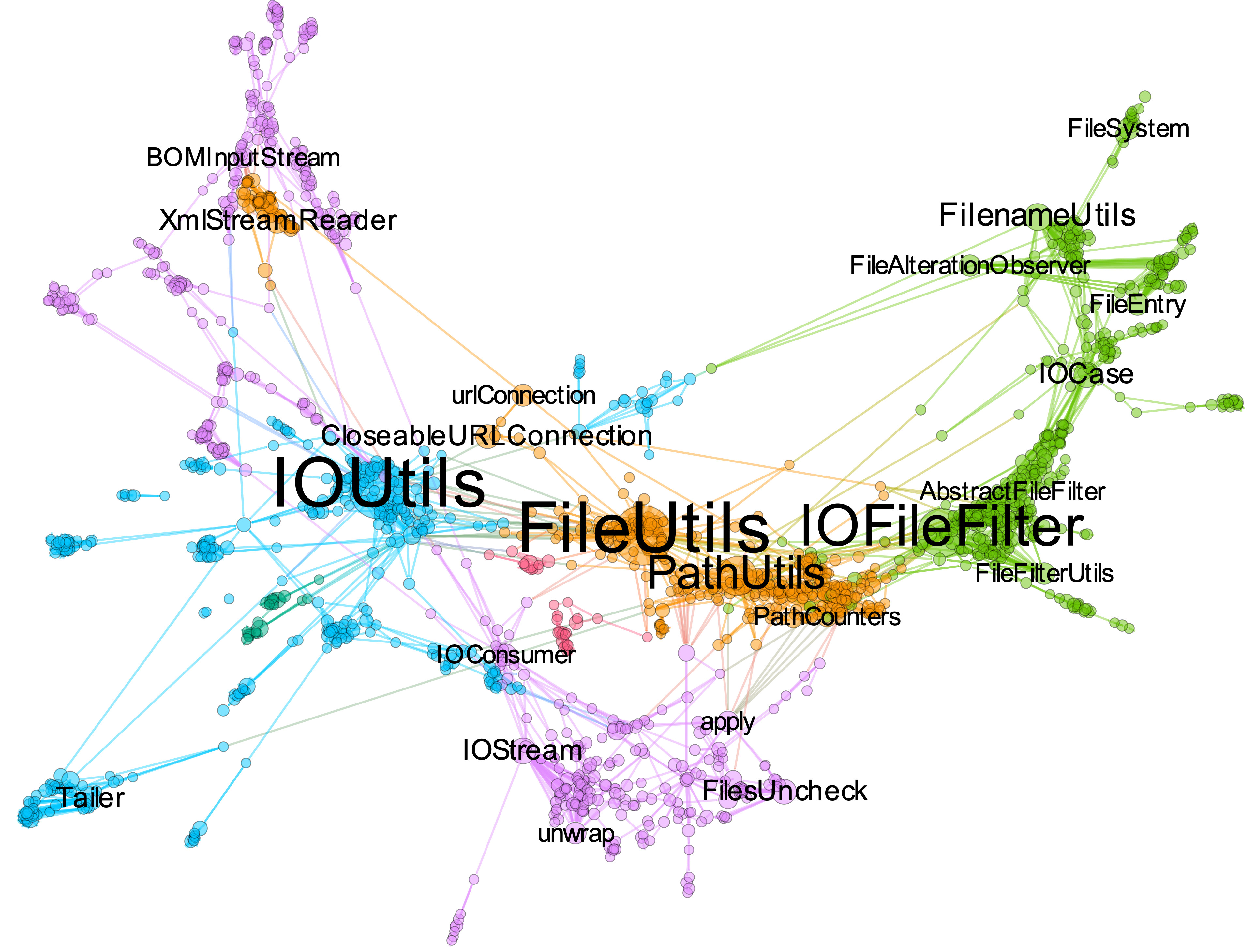}
    \caption{Top nodes for the \textit{commons-io} project. Colors denote partitions (modules) obtained from 6-partition split using the gpmetis algorithm. Visualized with Gephi.}
    \label{fig:gephi-partition-6}
\end{figure}

We can visualize the results of partitioning with the Gephi tool, by exporting the partitioning results as a graph, with additional node properties concerning partition number in a given split, in the graphml format: \lstinline{scg-cli -o gml partition commons-io 10}. The node colors presented in Fig. \ref{fig:gephi-partition-6} denote the assigned partitions.

\section{Impact}

With the enormous amount of new code created every day, speeding up software comprehension becomes ever more important. The \textit{scg-cli} tool, with the underlying theoretical concept of the Semantic Code Graph, enables comprehensive project analysis focused on understanding the code structure. The tool allows for quick insight into the project using a simple command line interface. Presenting project overview, finding critical code entities, and suggesting software partitioning (discovering modules) gives the programmer a good basic understanding into the project.

However, the capabilities of the tool do not end there. The semantic data describing the code structure is written in an open intermediate protobuf format, easy to consume by other tools for more in-depth analysis. To our best knowledge, no existing tool has this capability. Consequently, our approach can potentially impact future tools focusing on code dependency and structure analysis, and overall software comprehension. The \textit{scg-cli} tool itself provides functionality to export and load the SCG data into the Jupyter notebook, enabling a~split-apply-combine style of analysis using dataframes, a~well-established data analytics approach \cite{Wickham2011split}.

The \textit{scg-cli} is open sourced on the Github platform allowing further project collaborative development. Other researchers can contribute by adding SCG data generation capabilities for new languages, extending currently available set of functionalities, or adding more scripts into Jupyter notebooks to enable more detailed analysis. Hopefully, this software will also have a positive effect towards new generation of development tools focused on supporting software comprehension.

\section{Conclusions}

We presented \textit{scg-cli}, a command line tool supporting software comprehension of Java and Scala projects. The process of analysis begins with extracting semantic information from the project source code in the form of the Semantic Code Graph, written in an open protobuf-based format. The \textit{scg-cli} provides insight into project structure and basic metrics, finds crucial software entities, and helps discover modules by computing graph partitionings. More in-depth analysis can be performed using the export functionality, which includes a~Jupyter notebook environment.

In the future we would like to add support for additional languages, implement new functionalities, and build a~community around the \textit{scg-cli} Github project. This will allow us to get the feedback and the support to extend tool capabilities.

\section*{Acknowledgements}
The work of Krzysztof Borowski was supported in part by the AGH University of Krakow and the "Doktorat Wdrozeniowy" program of the Polish Ministry of Science and Higher Education, and by the VirtusLab company.

\bibliographystyle{plain} 
\bibliography{bibliography}

\end{document}
\endinput